\documentclass[aps, prd, twocolumn,superscriptaddress,longbibliography,numerical]{revtex4-2}

\usepackage{dsfont}
\usepackage{amsmath}
\usepackage{epsfig}
\usepackage{graphicx}
\usepackage{bm}
\usepackage{amssymb}
\usepackage{slashed}
\usepackage{xcolor}
\usepackage{inputenc}
\usepackage{calrsfs}
\usepackage[colorlinks,citecolor=blue,linkcolor=blue,urlcolor=blue,plainpages=false,pdfpagelabels]{hyperref}

\newcommand{\be}{\begin{equation}}
\newcommand{\ee}{\end{equation}}

\begin{document}
\title{Electron-phonon coupling in copper intercalated \textrm{Bi$_{2}$Se$_{3}$}}

\author{Maciej Wiesner}
\affiliation{Faculty of Physics, Adam Mickiewicz University, Poland}

\author{Kristie Koski}
\affiliation{Department of Chemistry, University of California Davis, Davis California USA}

\author{Antti Laitinen$^\dagger$}
\affiliation{Low Temperature Laboratory, Department of Applied Physics, Aalto University, P.O. Box 15100, 00076 Aalto, Finland}
\affiliation{Department of Physics, Harvard University, Cambridge, MA 02138, USA}
\author{Juuso Manninen}
\affiliation{Low Temperature Laboratory, Department of Applied Physics, Aalto University, P.O. Box 15100, 00076 Aalto, Finland}

\author{Alexander A. Zyuzin}
\affiliation{Low Temperature Laboratory, Department of Applied Physics, Aalto University, P.O. Box 15100, 00076 Aalto, Finland}
\affiliation{A. F. Ioffe Physical–Technical Institute, 194021 St. Petersburg, Russia}

\author{Pertti Hakonen}
\email{pertti.hakonen@aalto.fi}
\affiliation{Low Temperature Laboratory, Department of Applied Physics, Aalto University, P.O. Box 15100, 00076 Aalto, Finland}

\begin{abstract}

We report charge and heat transport studies in copper-intercalated topological insulator Bi$_2$Se$_3$ hybrid devices. Measured conductivity shows impact of quantum corrections, electron-electron and 
electron-phonon interactions. Our shot noise measurements reveal that heat flux displays a crossover between $T^2$ and $T^4$ with the increase of temperature. 
The results can be explained by a model of inelastic electron scattering on disorder, increasing the role of  transverse acoustic phonons in the electron-phonon coupling process.
\end{abstract}

\maketitle

\section{Introduction}
Charge and heat transport in metals are governed by the interplay between the electron scattering from impurities, quantum corrections, electron-electron (EE) and electron-phonon (EP) interactions. 
Here we study the temperature dependence of conductivity and heat flux in weakly copper-intercalated topological insulator (TI) \textrm{Bi$_{2}$Se$_{3}$}. 
Although, the quantum and EE corrections to the conductivity in \textrm{Bi$_{2}$Se$_{3}$} have been extensively studied \cite{Checkelsky2009,Checkelsky2011,Hong2011,Liu2012,Steinberg2011,Zhang2010,Stephen2020} and the strength of the electron-phonon coupling has been determined e.g. by using ARPES \cite{Pan2012} and surface phonons \cite{Zhu2012}, very little is known about the effect of the interference between EP and electron-impurity scattering on electrical transport, which has been of great interest in disordered metallic conductors \cite{Ptitsina1997} and nanowires \cite{Hsu2012}.
In disordered materials, for example intercalated crystals, the EP and electron-impurity interference processes result in a $T^2$ contribution to  conductivity \cite{Reizer_Sergeev}.

In general, the contribution of phonons to the low-temperature conductivity can be analysed by introducing two characteristic temperatures. First is the Bloch-Gr\"uneisen temperature $T_{\mathrm{BG}}$, which is associated with the particles' Fermi surface. Second is the temperature at which the phonon wave length becomes comparable to the mean free path of electrons due to scattering on impurities \cite{Reizer_Sergeev, Reizer_Sergeev_2, Sergeev2000, Chen2012}. It can be estimated as
$
T_{\mathrm{dis}}= 2\pi \hbar v_\mathrm{s}/(k_\mathrm{B} L_\mathrm{e})
$, 
where $v_\mathrm{s}$ is the sound velocity, $L_\mathrm{e}$ the electron mean-free path, and $\hbar, k_\mathrm{B}$ are the Planck and Boltzmann constants. Note the sound velocities of longitudinal and transverse modes are different, which has to be taken into consideration in an analysis of the EP interaction and determination of characteristic temperatures in the material.

Coupling of electrons to longitudinal and transverse acoustic phonons is a complex problem, which has been widely analysed in clean systems $T_{\mathrm{dis}}\ll T$ \cite{Mahan1990}. In pure materials, electrons do not interact with transverse phonons in the lowest order in the EP interaction term. Fundamentally, for clean \textrm{Bi$_{2}$Se$_{3}$}, the phonon contribution to the resistance in the bulk is given by the Bloch-Gr\"uneisen term $\propto T^5$ at low temperatures $T_{\mathrm{dis}}\ll T< T_{\mathrm{BG}}$ \cite{Giraud2011}, while $\propto T^4$ is expected for thin crystals \cite{Giraud2012}. 

In impure materials, however, electron interactions with transverse phonons contribute to the charge transport. In the case when the EP coupling is determined by the interference between phonons and disorder, the resistance at temperatures $T_{\mathrm{dis}} \lesssim T \lesssim T_{\mathrm{BG}}$ is given by \cite{Reizer_Sergeev, Ptitsina1997}:
\begin{equation} \label{GrindEQ__8_}
\frac{{R-R}_0}{R_0}=  \left[1-\frac{1}{2}{\left(\frac{v_{\mathrm{t}}}{v_{\ell}}\right)}^3 \left(1-\frac{\pi^2}{16}\right)\right] \frac{\beta_{\mathrm{t}} }{v_{\mathrm{t}}} \frac{(2\pi k_{\mathrm{B}}T)^2}{3 E_{\mathrm{F}} p_{\mathrm{F}}  \hbar}
\end{equation}
where $R_{0}$ is the resistance determined by the electron scattering on disorder,
$E_{\mathrm{F}}$ is the chemical potential, $p_{\mathrm{F}}$ is the Fermi momentum, $\beta_{\ell}$ and $\beta_{\mathrm{t}}$ are the constants of interaction with longitudinal
and transverse phonons with velocities $v_{\ell}$ and $v_{\mathrm{t}}$, respectively. The interaction constants satisfy ${\beta}_{\mathrm{t}}/{\beta}_{\ell }={\left(v_{\ell }/v_{\mathrm{t}}\right)}^2$ and can be retrieved from the experimental data. 
For \textrm{Bi$_{2}$Se$_{3}$}, taking into account $v_{\mathrm{t}}=1.7$\,km/s and $v_{\ell}=3.5$\,km/s , $\beta_{\mathrm{t}}/{\beta}_{\ell} \approx 4$ emphasises stronger contribution of transverse than longitudinal phonons, quite similar to disordered metal samples in Ref. \onlinecite{Ptitsina1997}.
We note that the crossover between $T^2$ and $T^5$ laws in resistivity depends on the mean free path and the coefficients of electron interaction with transverse and longitudinal phonons \cite{Sergeev2000}. 

The heat transport by electrons as opposed to relaxation by coupling to phonons can be distinguished in shot noise measurements. 
In particular, the heat flux between hot electrons and phonons with temperatures $T_{\mathrm{e}}$ and $T_{\mathrm{ph}}$, respectively, scales as
$
P\left(T_\mathrm{e},T_{\mathrm{ph}}\right) \propto T_{\mathrm{e}}^k-T_{\mathrm{ph}}^k
$
($T_{\mathrm{e}}$ was defined in Methods section). 

The power law $k$ is sensitive to the dimensionality of the sample \cite{Anghel2019}, disorder \cite{Reizer_Sergeev,Chen2012,Xu2016}, type of phonons \cite{Reizer_Sergeev}, screening \cite{Kubakaddi2007} and chirality \cite{Ansari2017} of charge carriers.
Here we distinguish the heat flux mechanism, which is determined by the transverse phonons at $T_{\mathrm{dis}} \lesssim T \lesssim T_{\mathrm{BG}}$ \cite{Sergeev2000}, 
\begin{equation} \label{eq:power3D}
P\left(T_{\mathrm{e}},T_{\mathrm{ph}}\right)=\frac{\pi^2 \beta_{\mathrm{t}}}{10 \hbar^3}\frac{V}{L_{\mathrm{e}} v_{\mathrm{F}} v_{\mathrm{t}} }\left[(k_{\mathrm{B}}T_{\mathrm{e}})^4-(k_{\mathrm{B}}T_{\mathrm{ph}})^4\right],
\end{equation}
where 
$v_{\mathrm{F}}$ is the Fermi velocity and $V$ is the volume of the sample.

In this paper we present experimental results for the temperature dependence of the conductivity and heat flux in the copper intercalated $\textrm{Bi}_2\textrm{Se}_3$. 
We focus on the signatures of electron-phonon interaction contribution to the observables.
First we comment on a crossover between EE and EP interaction corrections to the conductivity. Then we analyze the contribution of the EP interaction to the heat flux.

\section{Results and discussion}
\subsection{Conductivity}
Measurements of magnetoconductivity were performed at temperature $T=100$ mK and magnetic field ranging from 0 up to 3.5 T, as shown in Fig. \ref{fig:MGres}. 
Fitting of the results was based on a modified  Hikami-Larkin-Nagaoka approach 
\cite{Hikami1980,Lu2011,Sasmal2021}.
  Besides the quantum corrections, we employed both linear ($\propto B$) and quadratic ($\propto B^2$) magnetoresistance components. Quadratic terms arise either from classical magnetoresistance or spin–orbit scattering. The origin of the linear magnetoresistance (LMR) $\propto B$ is not apparent but, indeed, it has been observed in topological insulators such as Bi$_2$Se$_3$ \cite{Tang2011,Zhang2011} and Bi$_2$Te$_3$ \cite{Qu2010}.
 
 The magnetoconductivity $\Delta \sigma =\sigma \left(B,T\right) - \sigma\left(T\right)$ for Bi$_2$Se$_3$ can be written in the form \cite{Lu2011,Lu2014,Sasmal2021}: 
%
\begin{figure}[t]
	\centering
\includegraphics[width=9cm]{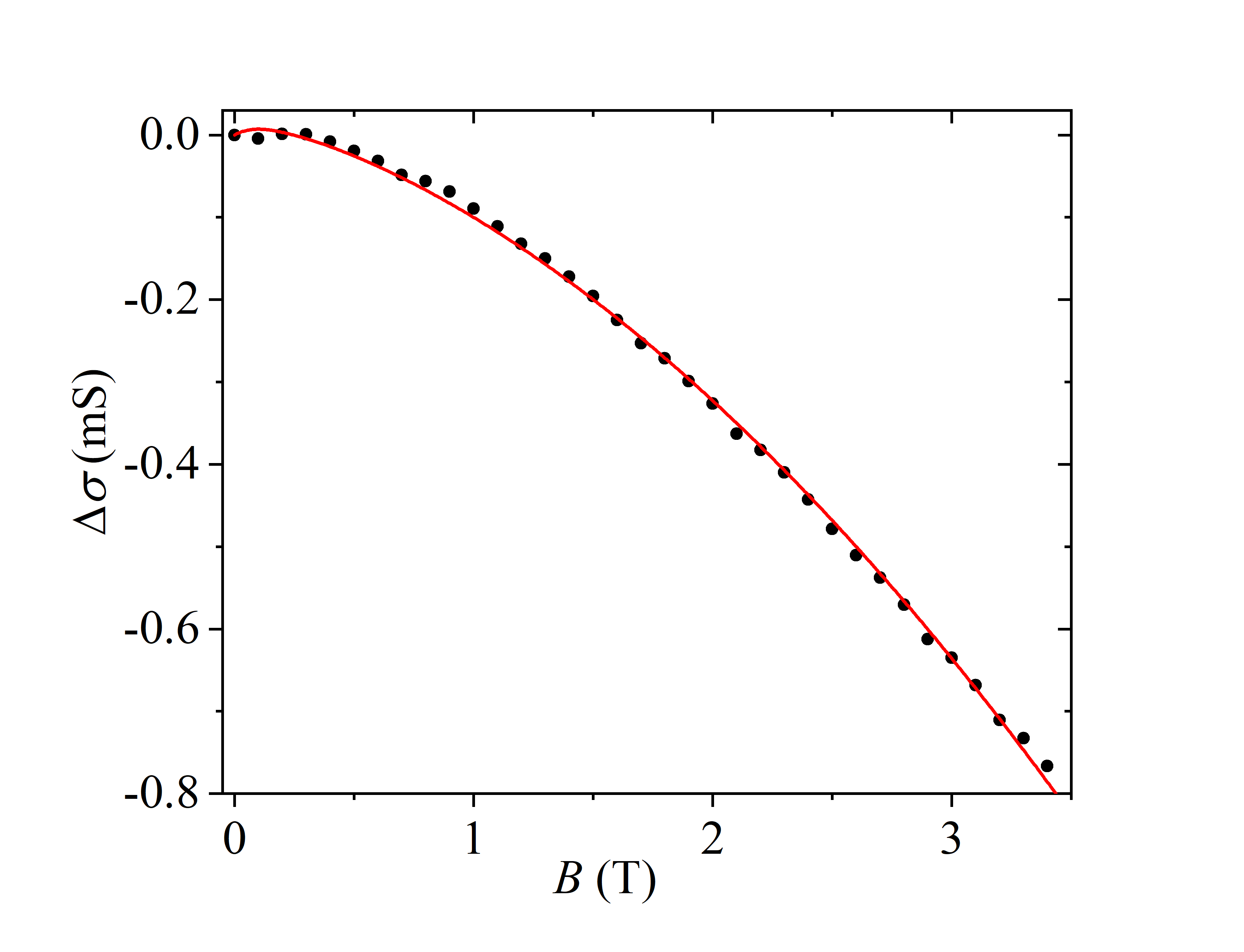} 
	\caption{Magnetoconductivity $\Delta \sigma =\sigma \left(B,T\right) - \sigma\left(T\right)$
	measured at $T=100$\,mK. The blue curve is a fit obtained using Eq. (\ref{eq:total MG}) together with a small linear magnetoresistance component. Quantum transport (WL) in the bulk is described by $\alpha=1$ and $L_{\mathrm{\phi}}=293$\,nm. Other fit parameters are discussed in the text. 
	\label{fig:MGres}}
\end{figure}
%
\begin{equation} 
\Delta \sigma_= 
 \frac{\alpha e^2}{2\pi^2\hbar }\left[\Psi\left(\frac{B_{\phi}}{B}+\frac{1}{2}\right)-\ln\left(\frac{B_{\phi}}{B}\right) \right] +\beta B^2 +\gamma B
  \label{eq:total MG}
\end{equation} 
where $\alpha$ denotes the strength and nature of quantum corrections and $B_{\mathrm{\phi}}=\frac{\hbar }{4e}L_{\mathrm{\phi}}^{-2}$ defines a characteristic field related to dephasing length $L_{\mathrm{\phi}}$. 
The term
$\beta B^2= - \sigma\left(T\right) {\mu}^2 B^2
- \frac{e^2}{24\pi^2\hbar } \left[{\left(\frac{B}{B_{\mathrm{SO}}+B_{\mathrm{e}}}\right)}^2 - \frac{3}{2 }{\left(\frac{B}{{\frac{4}{3}B}_{\mathrm{SO}}+B_{\phi}}\right)}^2 \right]$ where the first term $-\sigma\left(T\right) \mu^2 B^2$ denotes the classical, mobility-$\mu$-dependent magnetoresistance ($\mu^2 B^2 \ll 1$), while the latter terms are the quantum corrections specified using characteristics fields $B_{\mathrm{SO,e}}=\frac{\hbar }{4e}L_{\mathrm{SO, e}}^{-2}$ given by spin-orbit scattering length $L_{\mathrm{SO}}$ and the mean free path $L_{\mathrm{e}}$, respectively. 

In general, the parameter $\alpha$ is a measure of the relative strengths of the spin-orbit interaction, inter-valley scattering, magnetic scattering, and dephasing, as discussed in Ref.~\onlinecite{Garate2012} for topological insulators, and their value may vary between $-1 \dots +1$ (or even down to -2 \cite{Sasmal2021}). Using weak localization $\alpha=1$, our fit displayed in Fig. \ref{fig:MGres} yields  $L_{\mathrm{\phi}} = 293 \pm 100$ nm for the dephasing length, while the linear and quadratic terms amount to $\gamma = -0.11$ mS/T and $\beta=-41$\,$\mu$S/T$^2$, respectively. The value obtained for bulk dephasing length $L_{\mathrm{\phi}}$ is consistent with the earlier result of 300 nm obtained in Cu-doped Bi$_2$Se$_3$ \cite{Takagaki2012}. 

\begin{figure}[htp]
	\centering
\includegraphics[width=9.0cm]{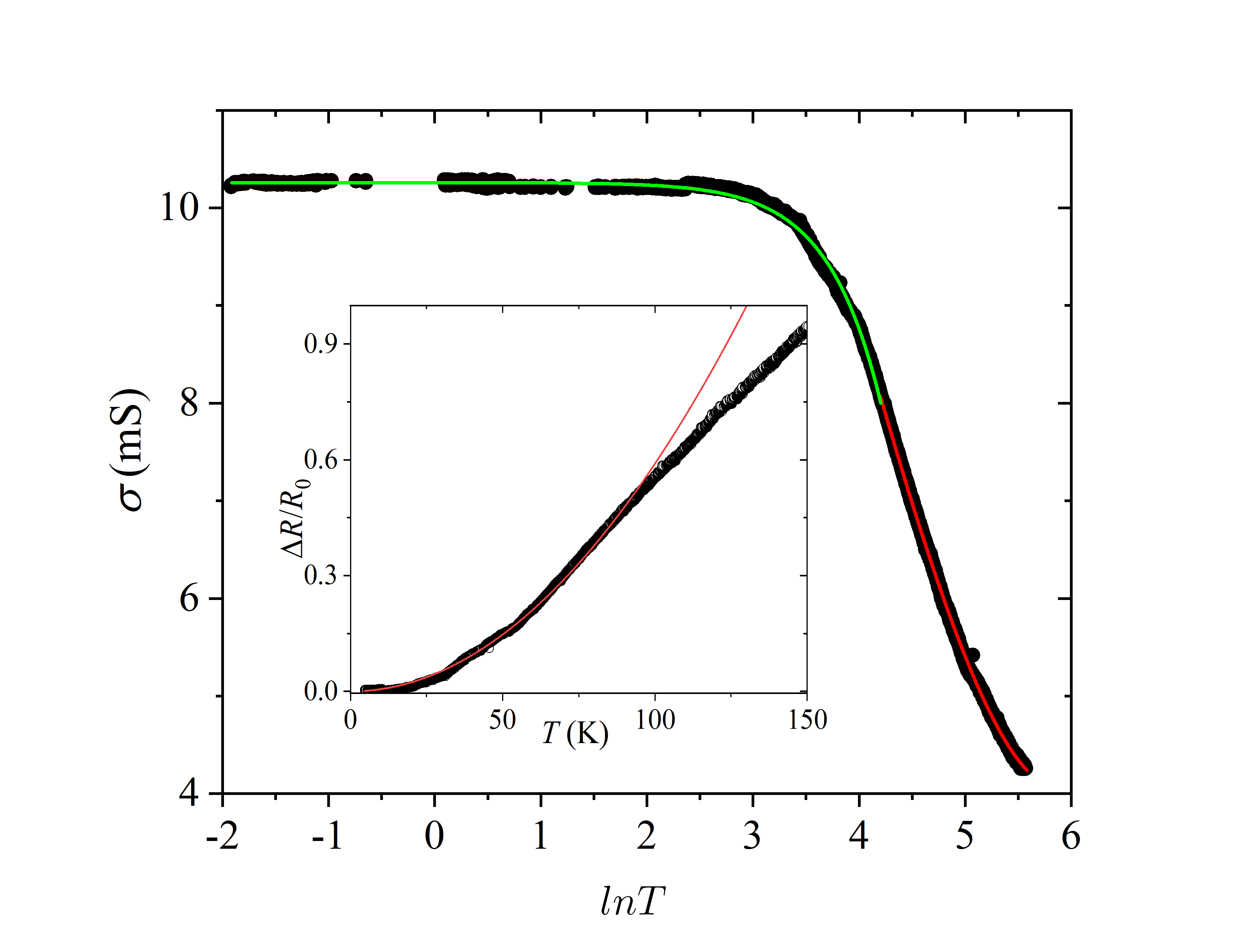} 
	\caption{ Temperature dependence of conductivity. The green line depicts a fit using Eq. \ref{GrindEQ__7_}, in which $\sigma_{0}=10$ mS, $k = 2$, $\alpha = 1$, and $d_2  \simeq -1 \times 10^{-6}$ S, while the red line illustrates linear resistivity at $T > T_{\mathrm{BG}}$ 
	The crossover temperature between the green and red curves is taken as the Bloch-Gr\"uneisen temperature $ T_{\mathrm{BG}}$.
	An inset shows relative change in the sheet resistance. The red curve is a fit to the data with $R_0$ subtracted using Eq. (\ref{GrindEQ__8_}) : $({R-R}_0)/R_0 = A T^2$, which gives for the prefactor $A = 59\times 10^{-6} $ K$^{-2}$. 
  \label{fig:dataB}}
\end{figure}

The transport in our sample is nearly fully governed by the bulk contribution. In this situation, the term $\beta B^2$ turns out to be dominated by classical magnetoresistance $\Delta R \propto \mu^2 B^2$, because $L_{SO}$ is expected to be much shorter than $L_{\phi} \simeq 300 $ nm in the bulk \cite{Steinberg2011} and $B_{SO} >> B_{\phi}$. Hence, we may identify $\beta = \sigma_0 \mu^2$ and obtain for the mobility of carriers $\mu=0.0657$ m$^2$/Vs. Using $g=\sigma_0\frac{ \pi \hbar}{e^2 }$,
where $\sigma_0=9.5$ mS, we estimate the mean free path $\lambda = g/k_F$=35\,nm, where the Fermi momentum $k_F=3.5\times 10^9$ 1/m was obtained from the Bloch-Gruneisen temperature (see text related to Fig. \ref{fig:dataB}).


Three principal remarks on the measured $\sigma(B)$ are in order.
First, neglecting contribution of the surface states, the bulk WL behavior in $\sigma(B)$ suggests a situation of weakly doped topological insulator with the chemical potential close to the conduction band gap \cite{Garate2012,Lu2014}.
To describe the data, the model of parabolic bulk conduction band may be utilized to describe the properties of copper intercalated Bi$_2$Se$_3$ topological insulator. In this limit, one may expect similar temperature dependencies of conductivity and heat flux between hot electrons and phonons.
Second, taking the bulk transverse and longitudinal phonon velocities we can estimate the characteristic temperatures at which thermal phonon wave-length becomes comparable to the electron mean free path. 
We estimate the $T_{\mathrm{dis}}= 2.3$\,K and $4.8$\,K for transverse and longitudinal phonons, respectively. 
Third, at low temperature $T \simeq 100$\,mK, the dephasing length $L_{\phi}=293$ nm is appreciably longer than $L_{\mathrm{e}}\simeq 35$ nm,  which  implies  several scattering events before the phase coherence is lost. The measured ratio\,\, $L_{\phi} / L_{\mathrm{e}} \sim 8$ implies that electrons in the bulk need several collisions to loose their phase coherence, so one can ascribe these collisions to weak scattering on disorder. Consequently, even weak phonon-induced scattering processes can become visible at low temperatures.

To proceed, the temperature dependence of conductivity was measured and the Bloch-Gr\"uneisen temperature was identified as an inflection point on the $R$ vs. $T$ dependence at $T_{\mathrm{BG}}=90~\textrm{K}$. We observe crossover between the logarithmic temperature dependence at low temperatures originating from competing WL, WAL, and EE interaction corrections and $T^2$ behavior at higher temperatures, which we assign to the EP interaction processes. We note that estimated $T_{dis}$ is slightly lower than the crossover temperature. The conductivity may be described by the formula:
\begin{equation}\label{GrindEQ__7_}
\sigma = \sigma_0+\frac{\alpha e^2}{2\pi^2\hbar } \ln (T/\mathrm{K})+d_k (T/\mathrm{K})^k,
\end{equation}
where $\alpha$ and $d_k$ are parameters and K stands for Kelvin. At $T \lesssim 10$K,  the temperature dependence of conductivity is governed by the quantum interference and EE interaction corrections for a system with effective 2D diffusion, which is described by the second term in Eq.  \ref{GrindEQ__7_}, having most commonly a positive sign due to the domination by EE interactions \cite{Lu2014}. 
In the interval $10 \mathrm{K} \lesssim T < T_{\mathrm{BG}}$, we observe signatures of the interference between the EP and electron-impurity scatterings Ref.~\cite{Reizer_Sergeev}. This mechanism is described by the third term in Eq. (\ref{GrindEQ__7_}), with $k = 2$. For surface states, the electron-phonon coupling has been found to display linear temperature dependence in resistance \cite{Kim2012}. Under large bias, coupling to optical phonons with exponential activation behavior has also been observed \cite{Costache2014}.

Equivalently, one can rewrite the last term in Eq. \ref{GrindEQ__7_} as a  correction to the resistance as is done in Eq. (\ref{GrindEQ__8_}). There the ratio of $v_{\ell}/v_{\mathrm{t}}$ determines the sign of the correction. Noting $\frac{1}{2}{\left(v_{\mathrm{t}}/v_{\ell}\right)}^3 \left(1- \pi^2/16\right) \approx 0.02$ suggests that the dominant contribution  to the EP interaction is due to transverse phonons in the investigated temperature range. 

To find the EP interaction constants, one can rewrite Eq. (\ref{GrindEQ__8_}) as $({R-R}_0)/R_0 = A T^2$. The prefactor $A = 59 \times 10^{-6}~\textrm{K}^{-2}$ is obtained by fitting the quadratic behavior to the experimental data in Fig. \ref{fig:dataB}. The Fermi momentum can be estimated as $k_{\mathrm{F}}= {k_{\mathrm{B}}T}_{\mathrm{BG}}/2\hbar v_{\mathrm{t}} \approx 3.5$ \textrm{nm$^{-1}$}, which allows us to estimate the chemical potential $E_{\mathrm{F}} = v_{\mathrm{F}} \hbar k_{\mathrm{F}}/2 \approx (0.46 \pm 0.11) ~\textrm{eV}$ for typical Fermi velocity $v_{\mathrm{F}} = (4 \pm 1) \times 10^{5}~\textrm{m/s}$ in \textrm{Bi$_{2}$Se$_{3}$}. Please note that the chemical potential is measured with respect to the bottom of the conduction band. We find the EP interaction factors $\beta_{\mathrm{t}} = 1.07 \pm 0.26$ and $\beta_{\ell}= \beta_{\mathrm{t}} \left( v_{\mathrm{t}}/v_{\ell}\right)^2 = 0.27\pm 0.07$.

\subsection{Heat flux between electrons and phonons}
At low temperatures, due to the depopulation of phonon modes and the degeneracy of the electron system, the two subsystems of a normal metal, the electrons and the phonons, become almost isolated from each other.
For this reason, they can `equilibrate' independently, that is, they may reach equilibrium distributions over the quasiparticle states at some effective temperatures that we shall denote by $T_{\mathrm{e}}$ (for electrons) 
and $T_{\mathrm{ph}}$ (for phonons).  Here we consider the so called ''hot electrons'' regime $T_{\mathrm{e}} \gg T_{\mathrm{ph}}$.
In the hot electron regime, electrons in Bi$_2$Se$_3$ have such a large energy that they can transfer heat by diffusion to the aluminum measurement leads effectively. Consequently, electronic thermal conductance in the Al lead dominates over heat transfer to phonons at bias voltages on the order of 1-2 mV \cite{Voutilainen2011}.

\begin{figure}[tb]
	\centering
\includegraphics[width=8.5cm]{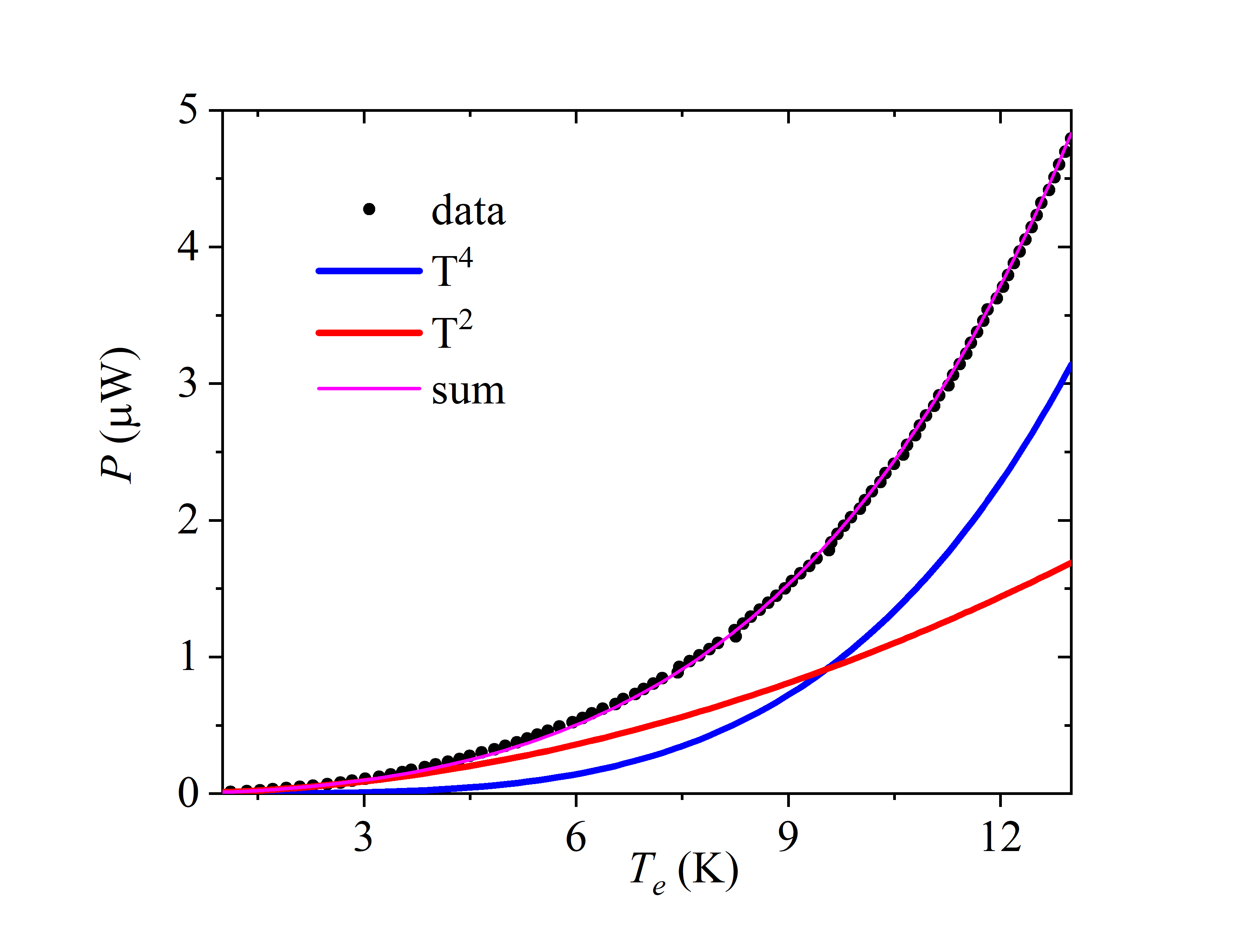} 
	\caption{ Joule heating power $P=IV$ as a function of temperature $T_e$ deduced from the measured shot noise. The power is fitted using $P(T_e)=c_2 T_e^2+c_4 T_e^4$ (magenta trace), which is a sum of electronic heat diffusion to the leads (red trace) and $P(T_e,T_{ph}=0) \propto T_e^4$  given in Eq. \ref{eq:power3D} (blue trace). The red and blue traces correspond to parameters $c_2=1.0\times10^{-8}$ $\mu$WK$^{-2}$  and $c_4=1.1\times10^{-10}$ $\mu$WK$^{-4}$, respectively. A crossover from electronic heat diffusion towards electron-phonon coupling dominated heat transfer is observed  at $T=9.4$\,K. Measurements were made at the phonon bath temperature of 50\,mK.} \label{fig:heatflow}
\end{figure}

The temperature dependence of the heat flow $P(T)$ out from electrons in Bi$_2$Se$_3$  displays both $T^2$ and $T^4$ dependencies as seen in Fig. \ref{fig:heatflow}, which depicts Joule heating $P=IV$ vs. $T_e$ determined from the measured non-equilibrium noise. 
At temperatures $T_e < 9~\textrm{K}$, electrons are the main heat carriers, which is reflected in the observed dependence $P\propto T_e^2$. 
Assuming homogeneous structure of investigated samples an analysis of the power dissipation at $T_e >  9$\,K can be performed taking $P\propto T_e^4$ (Fig. \ref{fig:heatflow}). In this case one can relate the energy dissipation with contribution of dynamic scattering potential to the electron interaction with transverse phonons as has been analysed in Refs. \onlinecite{Sergeev2000,Reizer_Sergeev_2}. A combined fit $P(T_e)=c_2 T_e^2+c_4 T_e^4$ yields a perfect match to the data using parameters given in the caption of Fig. \ref{fig:heatflow}

A $T^4$ temperature dependence could also arise for a heat flow limited by thermal boundary resistance in 3D systems (Kapitza resistance). It would reflect acoustic impedance
mismatch of phonons  across the interfaces made of Bi$_2$Se$_3$ and a SiO$_2$ substrate, as well as Bi$_2$Se$_3$ and metallic leads. This heat flow, $P({T_{ph}},{T_{0}}) = \frac{S}{4}{A_K}(T_{ph}^k - T_{0}^k)$, typically displays a Kapitza conductance of $A_K = 300 \dots 500$ Wm$^{-2}$K$^{-4}$ between metals and dielectric material \cite{Roukes1985,Swartz1989a,Elo2017}; here $T_{0}$ denotes the substrate phonon temperature. Using the contact area of the whole Bi$_2$Se$_3$ crystal to the SiO$_2$ substrate ($S_{\mathrm{SiO}_2}=4$ $\mu$m$^2$) and the Al leads ($S_{\mathrm{Al}}=3$ $\mu$m$^2$), we find that the Kapitza resistance cannot form a bottle neck in the thermal transport out from the sample.

An estimate for the strength of the interference contribution of the transverse phonons and random-scattering potential to the heat flux can be obtained from Eq. \ref{eq:power3D} in the limit $T_{\mathrm{ph}}=0~\textrm{K}$.
By using $\beta_{\mathrm{t}}=1.07$, $v_{\mathrm{F}} = 4\cdot10^{5}~\textrm{m/s}$, $L_{\mathrm{e}} =35~\textrm{nm}$, $v_{\mathrm{t}} =1.7~\textrm{km/s}$, and taking the volume of the sample as $V = 0.6\cdot 10^{-13}~\textrm{cm}^{3}$, Eq. \ref{eq:power3D} evaluated at $T_{\mathrm{e}} = 12~\textrm{K}$ yields for the heat flow $P(T_e,0) = 1.7~\mu \textrm{W}$, which is in pretty good agreement with the heat flow given by the $T^4$ component extracted from the experimental data in Fig. \ref{fig:heatflow}. Note that there is no explicit dependence of $P(T_e,0)$ on the Fermi velocity. However, if we identify $T_{BG}$ as 100 K, $P(12\, \mathrm{K},0)=2.2~\mu \textrm{W}$, nearly matching the measured power.


\section{Discussion and conclusions}
Typically,  charge and heat transport in disordered materials is considered as an effect of electron scattering from the static and vibrating potentials, taking into account that the vibrating potential is completely dragged by phonons. A tacit assumption is made about averaged velocity of the phonons in the investigated system. In many materials, including layered topological insulators, such assumption is invalid. Velocities of transverse and longitudinal phonons differ significantly. Measurements of charge transport at low temperatures provide clear distinction of the effect of scattering on transverse and longitudinal phonons. According to Sergeev and Mitin \cite{Sergeev2000} scattering of electrons on longitudinal phonons dominates in clean materials. Dynamic disorder, however, enhances electron coupling to transverse phonons. 

Our studies of the transport properties of copper -intercalated Bi$_{2}$Se$_{3}$ indicate that intercalation may provide a tool for modification of the charge and heat transport in a topological insulator. The first influence of intercalation is slight enhancement of linear magnetoresistance at low fields $B<2.5$T. 
Similar to the work of Ref. \onlinecite{Tang2011}, the observed LMR could be related to electronic inhomogeneity, enhanced by intercalation in our work. We believe that the intercalated amount $x=0.02$\% is too small to validate the use of the disorder-induced LMR model of Parish and Littlewood in our samples \cite{Parish2003}.


At  temperatures  $T\lesssim 10\mathrm{\,K}$, the quantum and EE interaction corrections dominate the conductivity and the electronic heat flux, yielding $\sigma \propto \ln (T)$ and $P \propto T^2$, respectively. 
We emphasize that $T_{\mathrm{dis}}$ in principle does not coincide with the crossover temperature between the electron-phonon and electron-electron mechanism contributions to the electric conductivity. Here, we estimate $T_{\mathrm{dis}}$ to be lower than the crossover temperature between $T^2$ and $T^4$ behaviors.
At $10\mathrm{\,K}\lesssim T \lesssim T_{\mathrm{BG}}$, the electron-transverse phonon interaction results in a $T^2$ temperature dependence of the conductivity, while the electron-phonon heat flux becomes $P\propto T^4$. The  $T^4$ dependence for heat flux is predicted also for clean 2D conductors \cite{Chen2012}, and it has recently been observed e.g. in epigraphene \cite{Karimi2021}. In our case, however, the combination of $T^2$ dependent resistivity and $T^4$ dependent heat flow point strongly towards a disordered system with electron-phonon-impurity interference phenomena.

\section{METHODS}

Single crystals of Bi$_{2}$Se$_{3}$ were grown using the vapour-liquid-solid (VLS) method and intercalated with zero-valent copper atoms using wet chemical intercalation \cite{Koski2012a}. It was assumed that copper intercalation in Bi$_{2}$Se$_{3}$ introduces disorder, which affects charge transport by enhanced scattering in the TI.
The intercalation ratio $x=0.02$\% was determined from the energy-dispersive X-ray spectroscopy (EDX). Sound velocity, in plane, was investigated using the 90a-degree equal-angle scattering geometry of the Brillouin light spectroscopy (BLS) with a five pass scanning tandem Fabry-Perot interferometer at $\lambda = 532 nm$. Results of the BLS experiments revealed transverse phonon velocity of 1700 m/s and longitudinal phonon velocity of 3500 m/s. 

We investigated hybrid devices made of Bi$_{2}$Se$_{3}$ topological insulator on SiO$_2$/Si substrate. We concentrate on representative data obtained for a sample with total dimensions $\textrm{Thickness}\times \textrm{Width}\times \textrm{Length} = 60\,\textrm{nm} \times 1560\,\textrm{nm} \times 2500\,\textrm{nm}$, intercalated with 0.02\% of Cu. A central section of length $L=650$ nm and width $W=1560$ nm was contacted  using titanium/aluminium electrodes (5/45 nm). The resistance of the sample was 41\,$\Omega$ at 4\,K.

Our conductivity and current fluctuation experiments were performed on two pulse-tube-based dilution refrigerators operated around 0.015 K and 0.05K. For experimental details, we refer to Refs. \onlinecite{Danneau2008,Oksanen2014}.
To determine the EP interaction and its effect on heat transport in our hybrid device, we deduce the non-equilibrium electronic temperature from shot noise measurements \cite{Wu2010,Betz2013,Laitinen2014}. 
The Fano factor $F$ is defined as the ratio of the measured noise level $S_{\mathrm{I}}$ and the Poissonian noise $S_{\mathrm{P}}$, $F= S_{\mathrm{I}}/S_{\mathrm{P}}$. For diffusive electron transport the electronic temperature $T_{\mathrm{e}}$ 
is linearly dependent on $F$ and the bias voltage $V$, $T_{\mathrm{e}}=F e V /2k_{\mathrm{B}}$ \cite{Wu2010}. Equivalently, we can use a calibration constant $\mathcal{M}$ to obtain the electronic temperature of the investigated system, $T_{\mathrm{e}}=\mathcal{M} S\left(I\right)$. The constant was obtained as the ratio of the difference of the shot noise $\Delta S$ measured across two temperatures $T_{1}=900$ mK and $T_{2}=20$ mK: $\mathcal{M} = (T_2-T_1)/\Delta S$. 
As defined in this manner,  $T_{\mathrm{e}}$ denotes a spatial average over the temperature distribution in the Joule heated  regime where EP scattering is investigated.
 
\subsection*{Acknowledgements}
This work was supported by Aalto University AScI grant and the Academy of Finland projects 310086 (LTnoise) and 336813 (CoE Quantum Technology Finland). The research leading to these results has received funding from the European Union’s Horizon 
2020 Research and Innovation Programme, under Grant Agreement no 824109.
The work was partially sponsored by Polish National Centre of Science (NCN) grant 2015/17/B/ST3/02391.  The work of AZ is supported by the Academy of Finland Grant No. 308339. AZ is grateful to the hospitality of the Pirinem School of Theoretical Physics.

\end{document}